\input{aipcheck}

\documentclass[
    ,final            
  ]
  {aipproc}

\layoutstyle{6x9}

\begin{document}

\title{Heavy Flavor Production at PHENIX at RHIC}


\author{Xiaorong Wang for PHENIX collaboration}{
  address={New Mexico State University, Las Cruces, NM 88003,
  USA\\
  Hua-zhong Normal University, Wuhan 430079, P.R. China} }

\begin{abstract}

A study of heavy flavor production in different collision systems
in various kinematic regions presents an opportunity to probe cold
nuclear medium and hot dense matter effects. Results from the
PHENIX experiment on $J/\psi$ and open charm production in Au+Au
and Cu+Cu collisions at $\sqrt{s_{NN}}$ =200 GeV are presented.
The data show strong $J/\psi$ suppression in central AA
collisions, similar to NA50 results, and strong suppression in
high $p_T$ open charm production. The $J/\psi$ production in Au+Au
and d+Au collisions is compared to understand the cold nuclear
medium effects. The data show significant cold nuclear effects in
charm production in d+Au collisions at forward and backward
rapidity ranges.
\end{abstract}

\maketitle


\section{introduction}
Heavy flavor is a sensitive probe of the gluon structure function
in the nucleon and its modification in nuclear matter. Charm
quarks are believed to be mostly created from initial gluon fusion
in hadronic collisions. Since they are massive, heavy flavor
hadrons are proposed to be ideal probes to study the early stage
dynamics in heavy-ion collisions.

$J/\psi$ and open charm production are two of the most important
hard probes of the hot dense matter created in Au+Au collisions.
It is also necessary to understand their production in the cold
nuclear medium in d+Au collisions as a reference in their
production in Au+Au collisions. Since the initial formation of
both open and closed charm is sensitive to initial gluon
densities, then gluon structure functions, shadowing or
anti-shadowing and initial state energy loss will all effect their
production. During the hadronization, the $J/\psi$ production
mechanism is different from that of open charm. In the final
state, the $J/\psi$ can be disassociated or absorbed, but for open
charm, the main nuclear medium effect is final state multiple
scattering and energy loss.

The PHENIX detector is well-suited for lepton measurements, At
mid-rapidity a Ring Image Cherenkov Detector (RICH) and an
electromagnetic calorimeters are used to identify electrons. The
PHENIX moun tracker and muon identifier work together to
reconstruct muons at forward and backward rapidities. One should
note here that we measure non-photonic electrons and prompt muons
to study open heavy flavor semi-leptonic decay, but we can not
separate charm and bottom contributions. The PHENIX beam-beam
counter provides a centrality and collision vertex measurement.

The PHENIX experiment\cite{phenix} has measured $J/\psi$ and open
charm production through observation of dilepton and semi-leptonic
decays respectively, in three rapidity ranges with the PHENIX muon
spectrometers and central arms, covering rapidity range of
$1.2<|\eta|<2.4$ and middle rapidity $|\eta| < 0.35$.

\section{$J/\psi$ production in d+Au and Au+Au collisions}

\begin{figure}
  \includegraphics[height=.32\textheight]{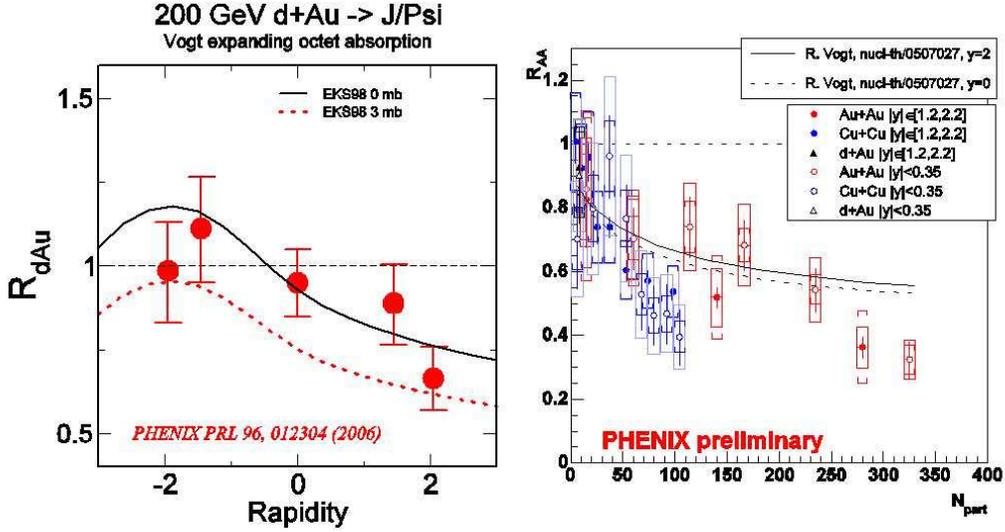}
  \caption{ a: Nuclear modification factor for d+Au collisions as a function of
  rapidity. Theory curves are from shadowing model EKS with the absorption cross section
  0mb(solid line) and 3mb(dotted line). b: Nuclear modification factor $R_{AA}$ as a function of
$N_{part}$ for Cu+Cu, Au+Au and d+Au. Theory curves show the same
model calculations as for d+Au with $\sigma_{ABS} =3mb$ in the y
=0 and y=2. }
\end{figure}

The nuclear modification factors is defined as the particle yield
per nucleon-nucleon collision relative to the yield in p+p
collisions. The PHENIX collaboration has measured the $J/\psi$
nuclear modification factor for d+Au \cite{jpsi038}, Cu+Cu and
Au+Au collisions. Figure 1a shows the nuclear modification factor
for d+Au collisions vs rapidity. A significant devaition from
unity is seen at the forward rapidity. Theory curves show the cold
nuclear medium calculations with shadowing and absorption present.
The current d+Au data probably only constrain absorption to
$\sigma_{ABS}\sim 0-3$ mb. Comparison with theoretical
calculations \cite{vogt} shows that a modest amount of absorption
with a modest shadowing scheme such as the EKS \cite{EKS}
prescription, can describe d+Au data.

In the hot dense matter in Au+Au collisions (Figure 1b, blue and
red solid squares), nuclear modification factor of $J/\psi$ shows
strong suppression at most central Au+Au collisions. Calculations
from same model \cite{vogtAuAu} to extrapolate such effects seen
in d+Au to Au+Au are presented on figure 1b. The curves show
$R_{AA}$ as a function of $N_{part}$. They account for shadowing
(following the EKS prescription) and nuclear absorption
$\sigma_{ABS}=3$mb (solid lines for y =0 and dashed lines for
y=2). The theoretical curves gives an idea of cold nuclear matter
effects on the $J/\psi$ production in A+A collisions. For the most
central Au+Au collisions measurements depart from the stronger
nuclear effects predictions (3 mb) suggesting effects beyond cold
nuclear matter are involved. For now, at least three classes of
models, Detailed transport\cite{DetailTrans}, sequential
melting\cite{SeqMelt} and recombination models
\cite{JpsiReco1}-\cite{JpsiReco3} exist that can roughly
accommodate the amount of anomalous suppression seen in our most
central preliminary results.

\section{open charm production in Au+Au and d+Au}

PHENIX studies of open heavy flavor are conducted via observation
of semi-leptonic decays. PHENIX has measured the $p_T$ spectra of
non-photonic electrons, $(e^+ + e^-)/2$ at mid-rapidity ($|\eta| <
0.35$) up to $p_T = 5$ GeV/c in d+Au and in Au+Au
collisions\cite{charm_auau_cen}.

The non-photonic single electron production at middle rapidity
follows binary collision scaling within experimental uncertainty
in d+Au collisions (see Figure 2a). Figure 2b shows $R_{AuAu}$
from the most central collisions compared to theoretical
predictions. In order to describe the data, large values of the
transport coefficient or gluon density have been assumed. In
addition, the bottom quark to charm quark ratio is not well known.
To gain a better theoretical understanding of the strong medium
modification observed, the charm and bottom quark signals need to
be disentangled.

\begin{figure}
  \includegraphics[height=.32\textheight]{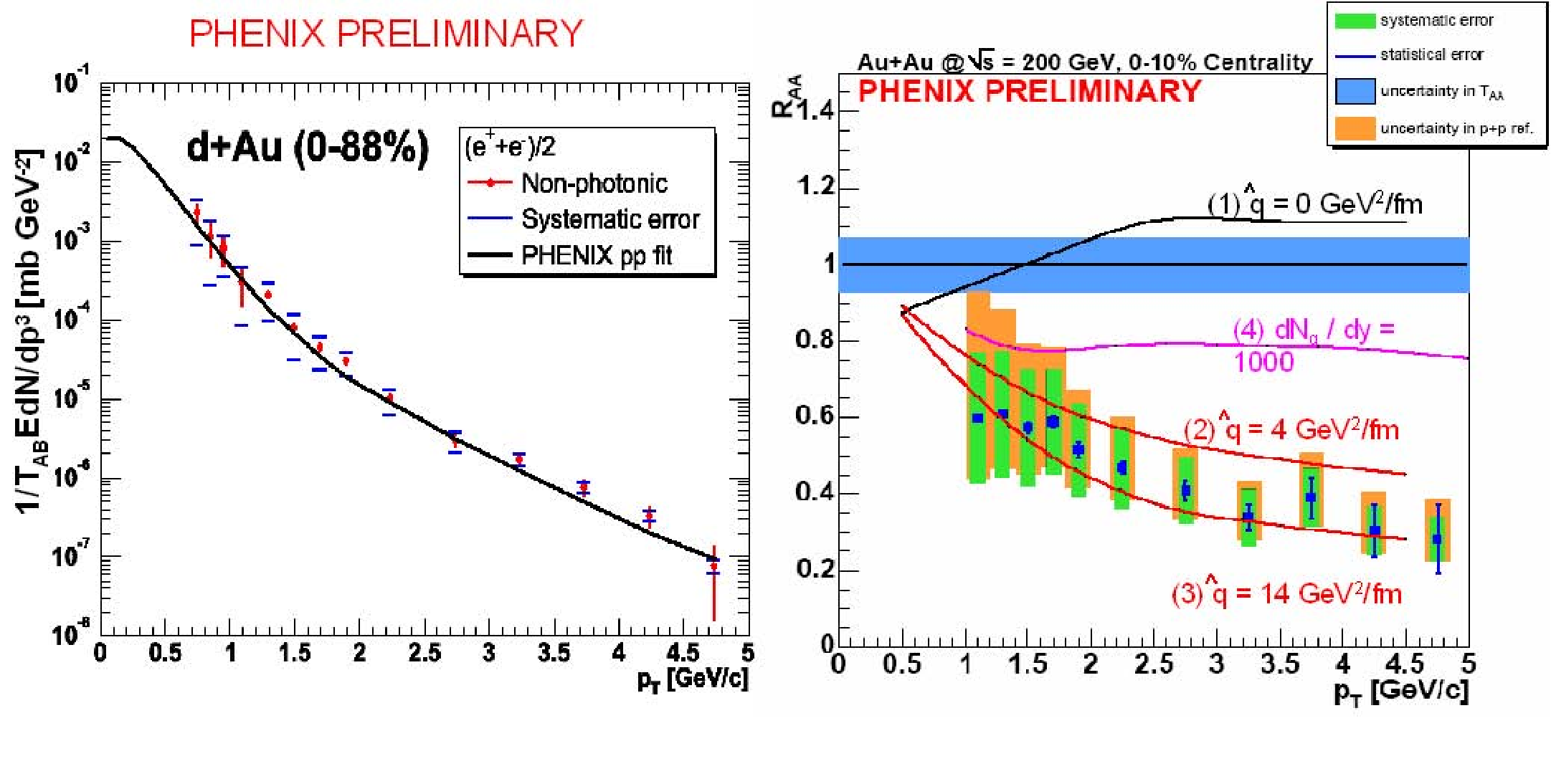}
  \caption{Open Charm production at y=0. a: Invariant cross section in dAu collisions overlayed with the $N_{coll}$ scaled pp reference.
  b:$R_{AA}$ for the 10{\%} most central Au+Au collisions to theoretical predictions. The three curves are taken from\cite{charm_raa_theory} ,
  with transport coefficients of 0, 4, and 14 GeV$^2$/fm, from top to bottom,
  and include only the charm contribution.}
\end{figure}

However in the forward and backward direction of d+Au collisions,
we probe different $x$ ranges which are sensitive to cold nuclear
effects such as parton shadowing, color glass condensate, initial
state energy loss, and coherent multiple scattering in final state
interactions. PHENIX has measured prompt single muon productions
in d+Au collisions.

Figure 3 shows the nuclear modification factor $R_{dAu}$ of prompt
muons at forward and backward rapidities. Suppression is observed
in the forward direction, which is consistent with CGC\cite{CGC}
and power correction pictures. The enhancement at the backward
direction needs more theoretical investigation as well as reduced
experimental errors. Anti-shadowing and recombination could lead
to such enhancement.

\begin{figure}
  \includegraphics[height=.3\textheight]{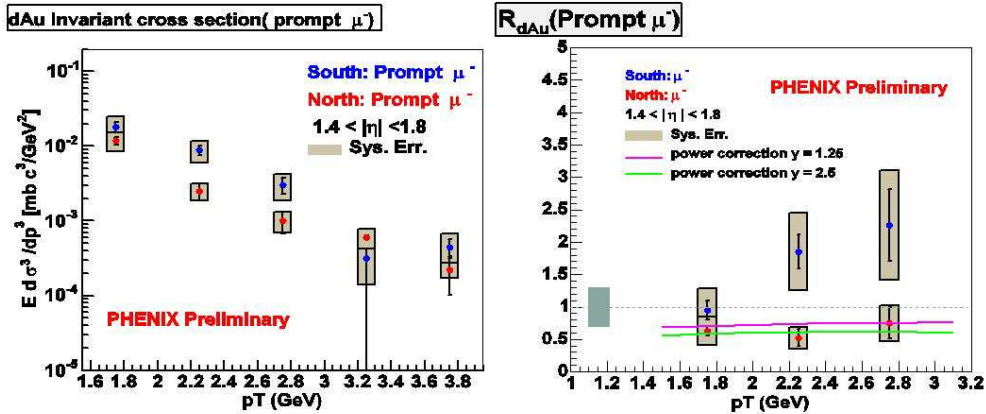}
  \caption{Invariant spectra(left) and nuclear modification factor of prompt muons (right) in
d+Au collisions. The theoretical curves are from a power
correction model at $\eta = 1.25$ and 2.5
\label{rdau_charm}\cite{ivan}. }
\end{figure}
\section{Summary}
PHENIX preliminary $J/\psi$ results show a large amount of
suppression, which is about a factor of three. However its
interpretation is not clear. First, the amount of normal, cold
nuclear matter suppression is not precisely defined and demands
better d+Au reference. Nevertheless, the observed suppression in
Au+Au seems to exceed the maximum suppression permitted by cold
nuclear matter effects. Three classes of models that raise the
number of surviving $J/\psi$s were presented. They all suppose the
formation of a QGP.

Open charm production in d+Au collisions in the forward and
backward directions suggests significant cold nuclear medium
effects. The open charm production in Au+Au collisions favors the
models with strong coupling and large energy loss. To understand
the energy loss mechanism, we need to separate the charm and
beauty contributions.

\end{document}